\documentclass[aps,prl,twocolumn,floatfix,nofootinbib,showpacs,superscriptaddress]{revtex4-1}

\usepackage{amsmath,amsfonts,amssymb,bm}
\usepackage{dsfont}
\usepackage{graphicx}
\usepackage{color}
\usepackage{soul}

\usepackage[mathscr,scaled=1.15]{urwchancal}
\DeclareFontFamily{OT1}{pzc}{}
\DeclareFontShape{OT1}{pzc}{m}{it}%
{<-> s * [1.15] pzcmi7t}{}
\DeclareMathAlphabet{\mathpzc}{OT1}{pzc}{m}{it}

\definecolor{purple}{rgb}{0.5,0,0.5}
\definecolor{blue}{rgb}{0.0,0,0.9}

\begin{document}

\title{Completing the picture of the Roper resonance}

\author{Jorge Segovia}
\affiliation{
Grupo de F\'{\i}sica Nuclear and Instituto Universitario de F\'{\i}sica Fundamental y Matem\'aticas (IUFFyM) Universidad de Salamanca, E-37008 Salamanca, Spain}

\author{Bruno El-Bennich}
\affiliation{
Laborat\'orio de F\'{\i}sica Te\'orica e Computacional, Universidade Cruzeiro do Sul, 01506-000 S\~ao Paulo, SP, Brazil}
\affiliation{Instituto de F\'{\i}sica Te\'orica, Universidade Estadual Paulista, 01140-070 S\~ao Paulo, SP, Brazil}

\author{Eduardo Rojas}
\affiliation{
Laborat\'orio de F\'{\i}sica Te\'orica e Computacional, Universidade Cruzeiro do Sul, 01506-000 S\~ao Paulo, SP, Brazil}
\affiliation{Instituto de F\'{\i}sica, Universidad de Antioquia, Calle 70 No. 52-21, Medell\'{\i}n, Colombia}

\author{Ian\;C.~Clo\"et}
\affiliation{Physics Division, Argonne National Laboratory, Argonne, Illinois 60439, USA}

\author{Craig\;D.~Roberts}
\affiliation{Physics Division, Argonne National Laboratory, Argonne, Illinois 60439, USA}

\author{Shu-Sheng Xu}
\affiliation{Department of Physics, Nanjing University, Nanjing 210093, China}

\author{Hong-Shi Zong}
\affiliation{Department of Physics, Nanjing University, Nanjing 210093, China}

\date{15 April 2015}

\begin{abstract}
We employ a continuum approach to the three valence-quark bound-state problem in relativistic quantum field theory to predict a range of properties of the proton's radial excitation and thereby unify them with those of numerous other hadrons.  Our analysis indicates that the nucleon's first radial excitation is the Roper resonance.  It consists of a core of three dressed-quarks, which expresses its valence-quark content and whose charge radius is 80\% larger than the proton analogue.  That core is complemented by a meson cloud, which reduces the observed Roper mass by roughly 20\%.  The meson cloud materially affects long-wavelength characteristics of the Roper electroproduction amplitudes but the quark core is revealed to probes with $Q^2 \gtrsim 3 m_N^2$.
\end{abstract}


\maketitle

%
\noindent\emph{1:Introduction}.\,---\,The strong-interaction sector of the Standard Model (SM) is thought to be described by quantum chromodynamics (QCD), a relativistic quantum field theory.  QCD is fascinating because it is plausibly a nonperturbatively well-defined quantum field theory \cite{millennium:2006}.  If so, then it is unique within the SM.  QCD is also distinguished by being formulated in terms of degrees-of-freedom -- gluons and quarks -- that are not readily accessible via experiment, \emph{i.e}.\ they are confined; and the forces responsible for this effect appear capable of generating more than 98\% of the mass of visible matter, in a process known as dynamical chiral symmetry breaking (DCSB) \cite{national2012Nuclear}.  With so much at stake, it is imperative to reveal and understand the measurable content of QCD.

Spectroscopy has long served as a valuable tool with which to reach such goals; and so it is with QCD.  The computation of the spectrum of hadrons, the collection of readily accessible states constituted from gluons and quarks, which was first given a semblance of order by the constituent-quark model \cite{GellMann:1964nj,Zweig:1981pd}, and subsequent comparison with modern experiment are an integral part of the international nuclear and particle physics effort.

Prominent amongst these investments are nucleon resonance ($N^\ast$) programmes, \emph{e.g}.\ at Jefferson Lab (JLab), at Bonn and Mainz, and in Japan \cite{Aznauryan:2011qj,Aznauryan:2012ba,Crede:2013kia,Mokeev:2013kka,Agashe:2014kda}, which seek answers to a range of critical questions, such as which three-quark states (baryons) and resonances are produced by QCD, and how are they constituted?  The accompanying theory effort is challenged by the fact that meaningful comparisons with the data are only possible within frameworks that: preserve the symmetries of QCD and the pattern by which they are broken; express the intrinsic mass-scale(s) and features associated with confinement and DCSB; and employ realistic kernels in baryon bound-state equations, which are necessarily relativistic.  Meeting these requirements calls for the development and application of nonperturbative methods in QCD.

An additional difficulty for both experiment and theory is that many excited states are short-lived and overlapping, \emph{i.e}.\ close in energy and broad in width.  This makes it hard to determine their quantum numbers and identify their production mechanisms.  A conspicuous case is the ``Roper resonance'', which has defied complete understanding for almost fifty years \cite{Roper:1964zza}.  The Roper (now designated \cite{Agashe:2014kda} $N(1440)\,{\textstyle \frac{1}{2}}^+$) is just like the proton, except for being 50\% heavier.  Until recently, it could not be explained from QCD by any theoretical method.  However, that has changed with the appearance \cite{Suzuki:2009nj} of a good theoretical case in support of the view that the Roper is the proton's first radial excitation, with its unexpectedly low mass arising from a dressed-quark core that is shielded by a meson cloud, which acts to diminish its mass \cite{national2012Nuclear}.

This pattern is repeated for many nucleon resonances \cite{Aznauryan:2008pe, Aznauryan:2009mx,Dugger:2009pn,Aznauryan:2009mx,Aznauryan:2011td,Mokeev:2012vsa}, with the cloud's impact apparently depending heavily on the state's quantum numbers \cite{Suzuki:2009nj,Aznauryan:2012ec}.  It is thus crucial to validate the proposed picture of the Roper.  That cannot be achieved by measurements of the mass and width alone, however.  One must also penetrate the meson cloud and thereby illuminate the putative dressed-quark core.  It should be possible to achieve this by measuring nucleon-resonance transition form factors in electroproduction experiments: whilst low-virtuality photons (total momentum $Q^2 \simeq 0$) are expected to be screened by the meson cloud, high virtuality photons ($Q^2 > m_N^2$, with $m_N$ the nucleon mass) may pierce the cloud, and thus can potentially expose the composition and distribution of the material within.  A chart of such electrocouplings for a large array of resonances could therefore provide a means by which to reveal the non-perturbative strong interaction phenomena that are essential to building $N^\ast$ states within the Standard Model.

Since the Roper has long resisted understanding, it has been a primary focus of the $N^\ast$ programme.  Experiments at JLab \cite{Aznauryan:2008pe, Dugger:2009pn, Aznauryan:2009mx, Aznauryan:2011td} have yielded a precise extraction of nucleon-Roper ($N\to R$) transition form factors and thereby exposed the first zero-crossing seen in any hadron form factor or transition amplitude.  It has also attracted a great deal of theoretical attention, \emph{e.g}.\ Refs.\,\cite{Cardarelli:1996vn, Glozman:1997ag, Capstick:1999qq, Tiator:2003uu, JuliaDiaz:2006av, Aznauryan:2007ja, Nagata:2008hi, deTeramond:2011qp, Edwards:2011jj, Lin:2011da, Wilson:2011aa, Aznauryan:2012ec, Engel:2013ig, Alexandrou:2013fsu, Kamleh:2013dga}.  Herein, via the first continuum treatment of this problem using the power of relativistic quantum field theory, we will show that all currently known features of the Roper resonance can be unified with those of the nucleon, $\Delta$-baryon (the lightest nucleon resonance, excited primarily via an $M1$ transition) and numerous other hadrons; and confirm the picture described above, \emph{viz}.\ the $N(1440)\,{\textstyle \frac{1}{2}}^+$ is the nucleon's first radial excitation.

\smallskip

\noindent\emph{2:Nucleon's first radial excitation}.\,---\,We compute the mass and wave function of the proton's first radial excitation using the Dyson-Schwinger equations (DSEs) \cite{Chang:2011vu, Bashir:2012fs, Cloet:2013jya}, an approach whose elements have an explicit connection with QCD.  In the limit of exact isospin symmetry, which is a good approximation within the strong interaction, the neutron and proton wave functions are indistinguishable, and the same is true for their excitations.

\begin{figure}[t]
\centerline{%
\includegraphics[clip,width=0.45\textwidth]{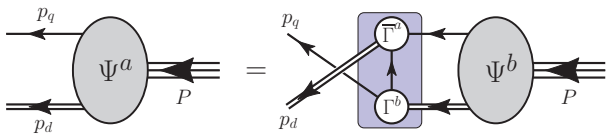}}
\caption{\label{figFaddeev} Poincar\'e covariant Faddeev equation.  $\Psi$ is the Faddeev amplitude for a baryon of total momentum $P= p_q + p_d$.  The shaded rectangle demarcates the kernel of the Faddeev equation: \emph{single line}, dressed-quark propagator; $\Gamma$,  diquark correlation amplitude; and \emph{double line}, diquark propagator.}
\end{figure}

The structure of a proton in relativistic quantum field theory is described by a Faddeev amplitude, obtained from a Poincar\'e-covariant Faddeev equation, which sums all possible quantum field theoretical exchanges and interactions that can take place between the three dressed-quarks that characterise its valence-quark content.
A dynamical prediction of Faddeev equation studies that employ realistic quark-quark interactions \cite{Qin:2011dd,Binosi:2014aea} is the appearance of nonpointlike quark$+$quark (diquark) correlations within baryons \cite{Cahill:1987qr, Cahill:1988dx, Maris:2002yu,Cloet:2008re,Eichmann:2009qa,Chang:2011tx,Cloet:2011qu}.  Consequently, the baryon bound-state problem is transformed into solving the linear, homogeneous matrix equation in Fig.\,\ref{figFaddeev}.

Empirical evidence in support of the presence of diquarks in the proton is accumulating \cite{Close:1988br,Cloet:2005pp,Wilson:2011aa,Cates:2011pz,Cloet:2012cy,Cloet:2014rja}.  It should be emphasised that these correlations are not the rudimentary, elementary diquarks introduced roughly fifty years ago in order to simplify treatment of the three-quark bound-state \cite{Lichtenberg:1967zz,Lichtenberg:1968zz}.  The two-body correlation predicted by modern Faddeev equation studies is not frozen; all dressed-quarks participate in all diquark clusters; and the baryon spectrum produced by this Faddeev equation has significant overlap with that of the three-quark constituent model and no simple relationship to that of the quark$+$elementary-diquark model.

\begin{figure}[t]
\centerline{%
\includegraphics[clip,width=0.72\linewidth]{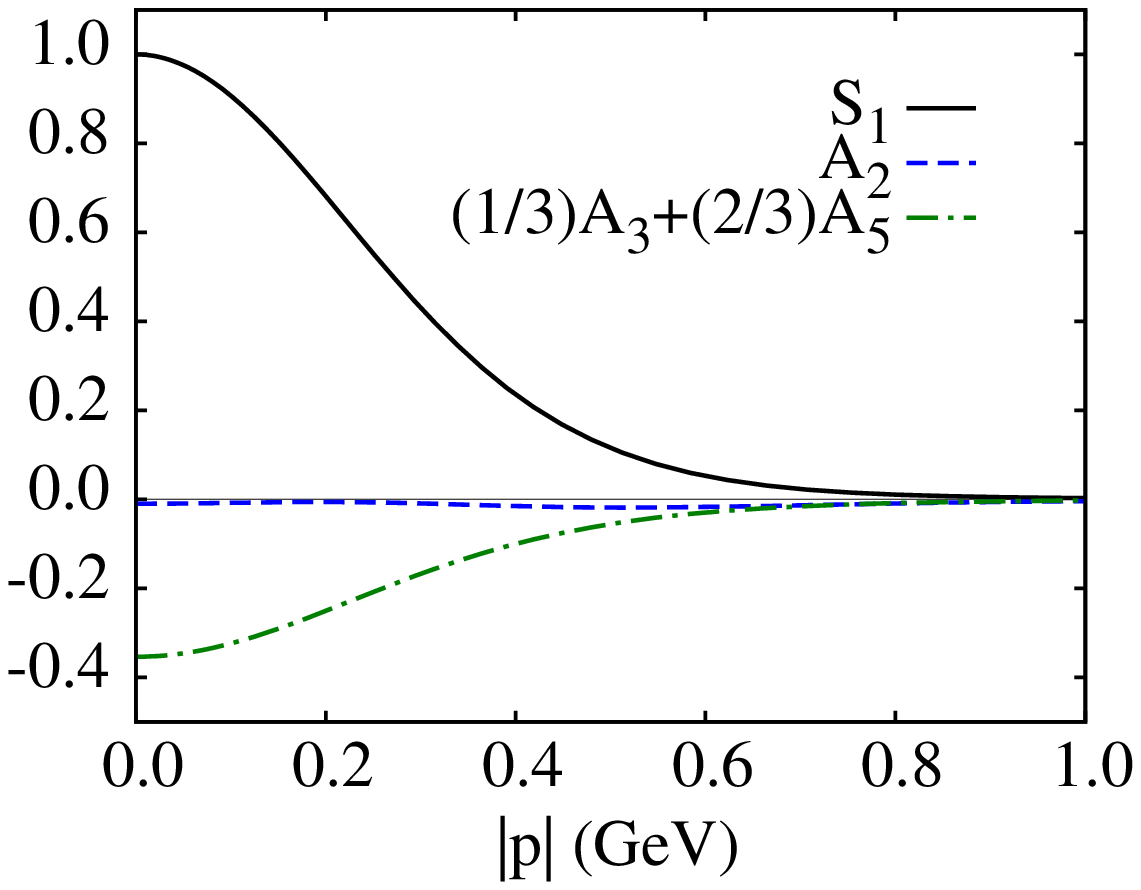}}
\centerline{%
\includegraphics[clip,width=0.72\linewidth]{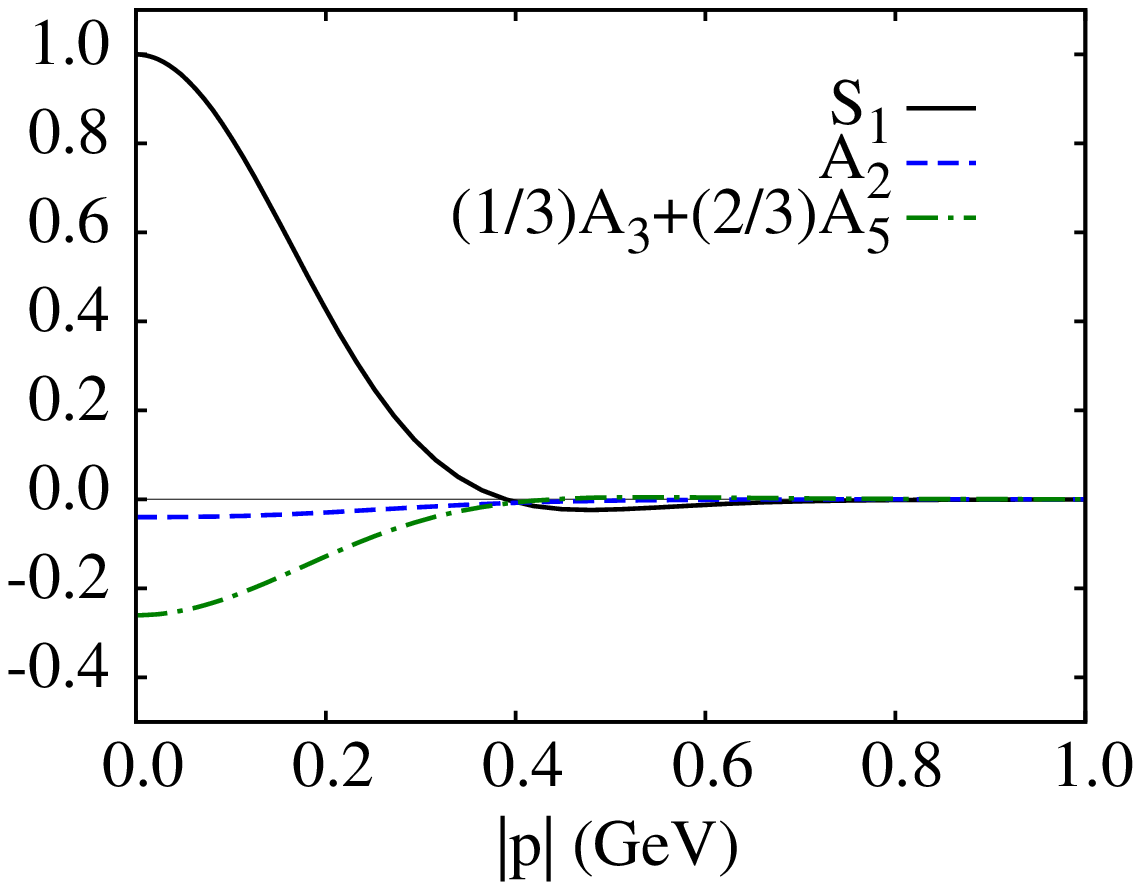}}
\caption{\label{figFA}
\emph{Upper panel}.  Zeroth Chebyshev moment of all $S$-wave components in the nucleon's Faddeev wave function, which is obtained from $\Psi$ in Fig.\,\ref{figFaddeev}, by reattaching the dressed-quark and -diquark legs.
\emph{Lower panel}. Kindred functions for the first excited state.
Legend: $S_1$ is associated with the baryon's scalar diquark; the other two curves are associated with the axial-vector diquark; and the normalisation is chosen such that $S_1(0)=1$.  Details are provided in Refs.\,\cite{Oettel:1998bk, Cloet:2007pi}.}
\end{figure}

Each element of the Faddeev equation depicted in Fig.\,\ref{figFaddeev} is specified in Ref.\,\cite{Segovia:2014aza}, which provides a successful description of the properties of the nucleon and $\Delta$-baryon, and is part of a body of work that unifies a large array of hadron properties (\emph{e.g}.\ see also Refs.\,\cite{Maris:1998hc,Maris:2003vk,Ivanov:2007cw,ElBennich:2010ha,ElBennich:2011py}).  With these inputs, we constructed the Faddeev equation kernel and used \emph{ARPACK} software \cite{Arpack} to obtain the mass and Faddeev amplitude of the nucleon and its first $J^P=1/2^+$ excited state.  The masses are (in GeV):
\begin{equation}
\label{eqMasses}
\mbox{nucleon\,(N)} = 1.18\,,\;
\mbox{nucleon-excited\,(R)} = 1.73\,.
\end{equation}
These values correspond to the locations of the two lowest-magnitude $J^P=1/2^+$ poles in the three-quark scattering problem.  The associated residues are the Faddeev wave functions, which depend upon $(\ell^2,\ell \cdot P)$, where $\ell$ is the quark-diquark relative momentum.  In Fig.\,\ref{figFA} we depict the zeroth Chebyshev moment of all $S$-wave components in that wave function, \emph{i.e}.\ projections of the form
\begin{equation}
{\mathpzc W}(\ell^2;P^2) = \frac{2}{\pi} \int_{-1}^1 \! dx\,\sqrt{1-x^2}\,
{\mathpzc W}(\ell^2,x; P^2)\,,
\end{equation}
where $x=\ell\cdot P/\sqrt{\ell^2 P^2}$.  Drawing upon experience with quantum mechanics and with excited-state mesons studied via the Bethe-Salpeter equation \cite{Holl:2004fr,Qin:2011xq,Rojas:2014aka}, the appearance of a single zero in $S$-wave components of the Faddeev wave function associated with the first excited state in the three dressed-quark scattering problem indicates that this state is a radial excitation.

Let us return to the masses in Eq.\,\eqref{eqMasses}.  The empirical values of the pole locations for the first two states in the nucleon channel are \cite{Agashe:2014kda,Suzuki:2009nj}: $0.939\,$GeV and $1.36 - i \, 0.091\,$GeV, respectively.  (The physical Roper is unstable and hence the associated pole has an imaginary part.)  At first glance, these values appear unrelated to those in Eq.\,\eqref{eqMasses}.  However, deeper consideration reveals \cite{Eichmann:2008ae,Eichmann:2008ef} that the kernel in Fig.\,\ref{figFaddeev} omits all those resonant contributions which may be associated with the meson-baryon final-state interactions that are resummed in dynamical coupled channels models \cite{Suzuki:2009nj,Kamano:2013iva,Doring:2014qaa} in order to transform a bare-baryon into the observed state.  Our Faddeev equation should therefore be understood as producing the dressed-quark core of the bound-state, not the completely-dressed and hence observable object.

For the nucleon itself, clothing the quark-core by including resonant contributions to the kernel produces a physical nucleon whose mass is approximately 0.2\,GeV lower than that of the dressed-quark core \cite{Ishii:1998tw,Hecht:2002ej}.  Similarly, clothing the $\Delta$-baryon's dressed-quark core lowers its mass by $\approx 0.16\,$GeV \cite{Suzuki:2009nj}.   It is therefore no coincidence that (in GeV) $1.18-0.2 = 0.98\approx 0.94$, \emph{i.e}.\ the nucleon mass in Eq.\,\eqref{eqMasses} is 0.2\,GeV greater than the empirical value.  A successful body of work on the baryon spectrum \cite{Chen:2012qr}, and nucleon and $\Delta$ elastic and transition form factors \cite{Cloet:2008re,Segovia:2014aza,Roberts:2015dea} has been built upon precisely this knowledge of the impact of omitting resonant contributions and the magnitude of their effects.

Crucial, therefore, is not a comparison between the empirical value of the Roper resonance pole-position and the computed quark-core mass of the nucleon's radial excitation but, instead, that between the quark-core mass and the value determined for the mass of the meson-undressed bare-Roper in Ref.\,\cite{Suzuki:2009nj}, \emph{viz}. (in GeV)
\begin{equation}
\label{eqMassesA}
\begin{array}{l|cc|c}
            & \mbox{R}_{{\rm core}}^{\rm herein}
            & \mbox{R}_{{\rm core}}^{\mbox{\footnotesize \cite{Wilson:2011aa}}}
            & \mbox{R}_{\rm bare}^{\mbox{\footnotesize \cite{Suzuki:2009nj}}} \\\hline
\mbox{mass} & 1.73 & 1.72 & 1.76
\end{array}\,.
\end{equation}
The bare Roper mass in Ref.\,\cite{Suzuki:2009nj} agrees with both our quark-core result and that obtained using a refined treatment of a vector$\,\otimes\,$vector contact-interaction \cite{Wilson:2011aa}.  This is notable because all these calculations are independent, with just one common feature; namely, an appreciation that measured hadrons can realistically be built from a dressed-quark core plus a meson-cloud.

\smallskip

\noindent\emph{3:Nucleon-Roper transition form factors}.\,---\,The agreement in Eq.\,\eqref{eqMassesA} is suggestive but not conclusive.  As observed in the Introduction, precise empirical information is available on the nucleon-Roper transition form factors.  Thus, if the picture we are describing is valid, then combining the solutions of the Faddeev equation in Fig.\ref{figFaddeev} for both the ground-state nucleon and its radial excitation should produce transition form factors that possess an understandable connection with available data and, indeed, match in accuracy the predictions for the nucleon and $\Delta$-baryon elastic and transition form factors obtained using the same approach \cite{Segovia:2014aza,Roberts:2015dea}.

In order to compute the electromagnetic $N\to R$ transition form factor, one must first calculate the analogous elastic form factors for the proton and its radial excitation because the $Q^2=0$ values of the associated charge form factors fix the normalisation of the transition.  Such calculations proceed from the Poincar\'e-covariant electromagnetic current for a spin-half baryon:
\begin{equation}
\textstyle
i e\,
\bar u_{f}(P_f)\big[ \gamma_\mu^T F_{1}^{fi}(Q^2)+\frac{1}{m_{{fi}}} \sigma_{\mu\nu} Q_\nu F_{2}^{fi}(Q^2)\big] u_{i}(P_i)\,,
\label{NRcurrents}
\end{equation}
where: $P_{i,f}$ are, respectively, the four-momenta of the incoming/outgoing baryon, each with mass $m_{i,f}$ so that $P_{i,f}^2=-m_{i,f}^2$; $Q=P_f-P_i$; $m_{{fi}} = (m_f+m_{i})$; and $\gamma^T \cdot Q= 0$.
In computing all form factors, we follow Refs.\,\cite{Cloet:2008re, Wilson:2011aa, Segovia:2014aza} in every respect, including formulation of the current \cite{Oettel:1999gc,Chang:2011tx}.  The transition form factors are obtained from the nucleon elastic form factor expressions by replacing all inputs connected with the final state by those for the radial excitation associated with the wave function in Fig.\,\ref{figFA}.  The critical issue is whether the form factors thus obtained have any relationship to those measured in the proton-Roper transition.

The QCD-based Faddeev equation predicts the existence of diquark correlations within baryons, so we first compare the diquark content of the nucleon and its radial excitation.  That information is contained in $F_1(Q^2=0)$, \emph{i.e}.\ the zero-momentum value of the elastic Dirac form factor \cite{Wilson:2011aa, Roberts:2013mja, Segovia:2014aza}; and we find:
\begin{equation}
\label{Pdiquark}
\begin{array}{l|cc|c}
        & N    & R    & N_{U}\\\hline
P_{J=0} & 62\% & 62\% & 30\% \\
P_{J=1} & 38\% & 38\% & 70\% \\
\end{array}\,;
\end{equation}
namely, the relative strength of scalar and axial-vector diquark correlations in the nucleon and its radial excitation is the same.  The last column in Eq.\,\eqref{Pdiquark} reports the diquark content of the unphysical state corresponding to the largest eigenvalue $(\lambda > 1)$ of the Faddeev kernel at $M=1.73\,$GeV: in this ``off-shell nucleon'', the diquark content is orthogonal to that of the on-shell $(\lambda=1)$ radial excitation, as it should be.  (See Ref.\,\cite{Krassnigg:2003wy} for details concerning the $P^2$-dependent character of the eigenvalue spectrum of a Poincar\'e-covariant bound-state kernel.)

The prediction $P_{J=0,1}^R\approx P_{J=0,1}^N$ contrasts starkly with the contact-interaction result \cite{Wilson:2011aa,Chen:2012qr}: $P_{J=0}\approx 0$.  In our view the latter should now be viewed as an artefact of the contact-interaction, which owes to the (over-)simplicity of the Faddeev kernel in that case.  This conforms with observations made elsewhere \cite{GutierrezGuerrero:2010md, Roberts:2011wy, Wilson:2011aa, Chen:2012txa}, \emph{viz}.\ that whilst a contact-interaction can yield useful insights into hadron static properties, it often fails in connection with structural properties that probe energy scales in excess of that associated with DCSB \cite{Chang:2011vu, Bashir:2012fs, Cloet:2013jya}: $M_D\simeq 0.4\,$GeV.

\begin{figure}[t]
\centerline{%
\includegraphics[clip,width=0.85\linewidth]{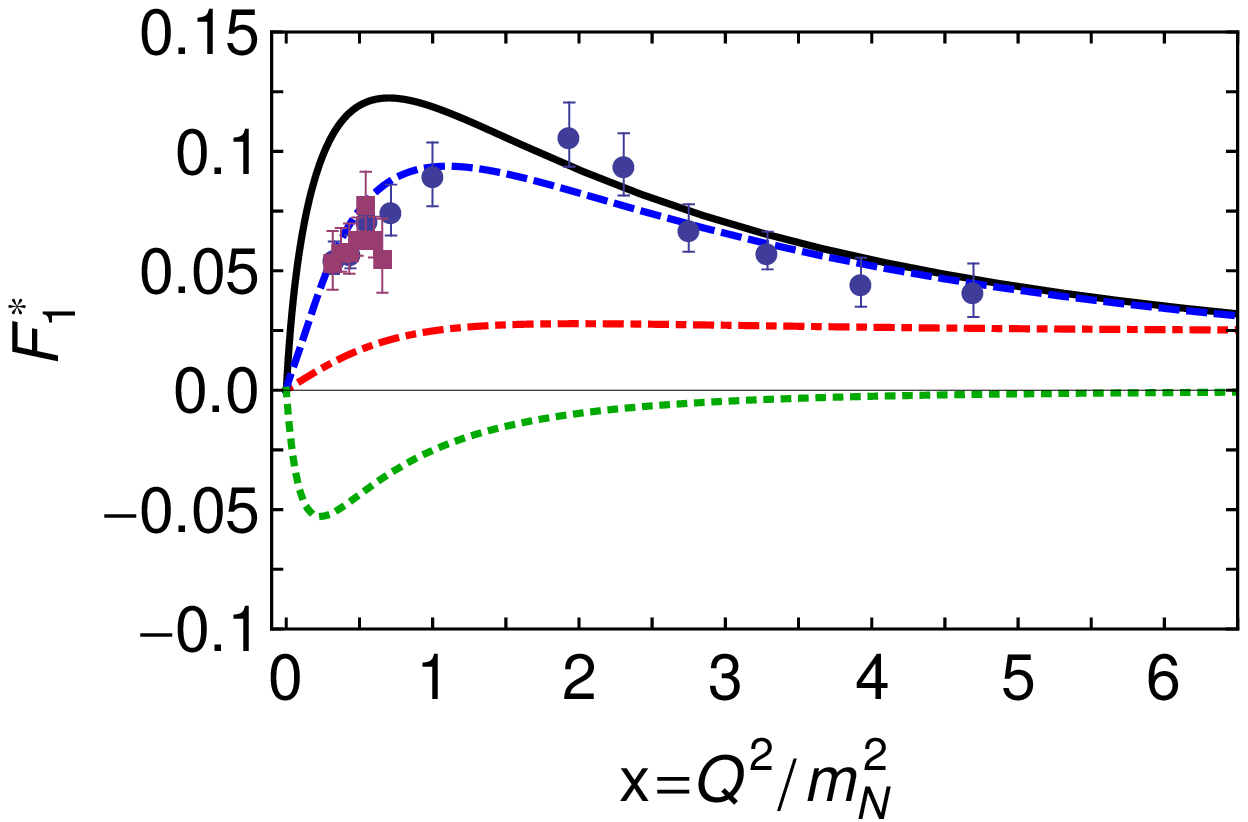}}
\centerline{%
\includegraphics[clip,width=0.85\linewidth]{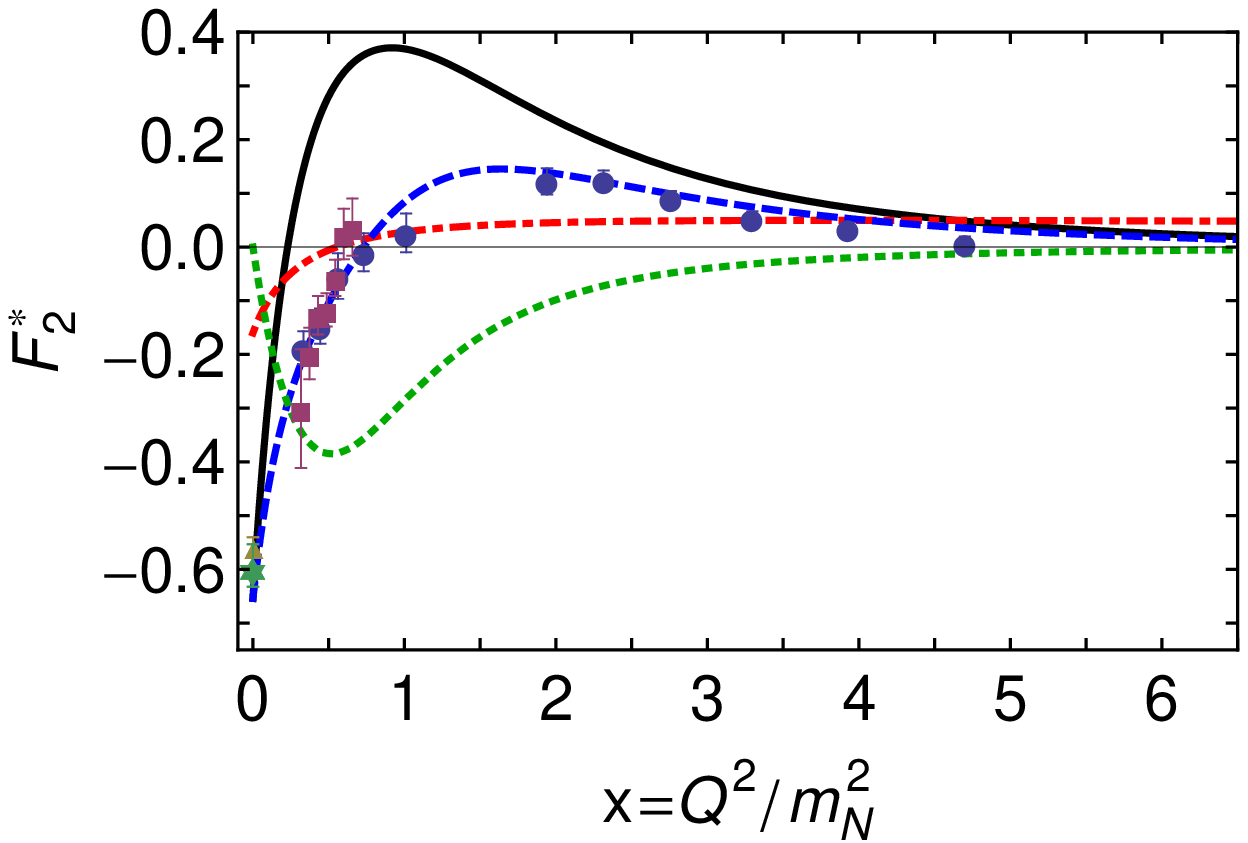}}
\caption{\label{figFT}
\emph{Upper panel} -- $F_{1}^{\ast}$ as a function of $x=Q^2/m_N^2$.  Solid (black) curve, our prediction; dot-dashed (red) curve, contact-interaction result \cite{Wilson:2011aa}; dotted (green) curve, inferred meson-cloud contribution; and dashed (blue) curve, anticipated complete result.
\emph{Lower panel} -- $F_{2}^{\ast}(x)$ with same legend.
Data in both panels: circles (blue) \cite{Aznauryan:2009mx};
triangle (gold) \cite{Dugger:2009pn};
squares (purple) \cite{Mokeev:2012vsa};
and star (green) \cite{Agashe:2014kda}.}
\end{figure}

Charge radii may also be computed from the elastic form factors; and we find $r^{ \Psi}_{R^+}/r_{p}^{\Psi}=1.8$, \emph{i.e}.\ a radius for the radial excitation that is 80\% larger than that of the ground-state.  The ratio of magnetic radii is $1.6$.

With the normalisations fixed, it is straightforward calculate to the $p\to R$ transition form factors.  Our results are displayed in Fig.\,\ref{figFT}.  The upper panel depicts the Dirac transition form factor $F_{1}^{\ast}=F_1^{Rp}$, which vanishes at $x=0$ owing to orthogonality between the proton and its radial excitation.  Our calculation agrees quantitatively in magnitude and qualitatively in trend with the data on $x\gtrsim 2$.  Nothing was tuned to achieve these results.  Instead, the nature of our prediction owes fundamentally to the QCD-derived momentum-dependence of the propagators and vertices employed in formulating the bound-state and scattering problems. This point is further highlighted by the contact-interaction result: with momentum-independent propagators and vertices, the prediction disagrees both quantitatively and qualitatively with the data.  Experiment is evidently a sensitive tool with which to chart the nature of the quark-quark interaction and hence discriminate between competing theoretical hypotheses; and it is plainly settling upon an interaction that produces the momentum-dependent dressed-quark mass which characterises QCD \cite{Bowman:2005vx, Bhagwat:2006tu, Roberts:2007ji}.

The mismatch between our prediction and the data on $x\lesssim 2$ is also revealing.  As seen previously, \emph{e.g}.\ Refs.\,\cite{Cloet:2008re, Segovia:2014aza, Roberts:2015dea}, this is the domain upon which meson-cloud contributions are expected to be important.  An estimate of that contribution is provided by the dotted (green) curve in Fig.\,\ref{figFT}.  If this curve is added to our prediction, then one obtains the dashed (blue) curve, which is a least-squares fit to the data on $x\in (0,5)$.  The correction curve has fallen to just 20\% of its maximum value by $x=2$ and vanishes rapidly thereafter so that our prediction alone remains as the explanation of the data.

The lower panel of Fig.\,\ref{figFT} depicts the Pauli form factor, $F_{2}^{\ast}=F_2^{Rp}$.  All observations made regarding $F_{1}^{\ast}$ also apply here, including those concerning the estimated meson-cloud contributions.  Importantly, the existence of a zero in $F_{2}^{\ast}$, a prominent feature of the data, is not influenced by meson-cloud effects, although its precise location is.  (The same is true of the $p\to\Delta^+$ electric transition form factor.)  Thus any realistic approach to the proton-Roper transition must describe a zero in $F_{2}^{\ast}$.
It is worth noting in addition that our prediction $F_{2}^{\ast}(x=0)=-0.65$, \emph{i.e}.\ for the Pauli form factor at the photoproduction point, is consistent with contemporary experiment: $-0.58 \pm 0.02$ \cite{Dugger:2009pn} and $-0.62 \pm 0.04$ \cite{Agashe:2014kda}.

\smallskip

\noindent\emph{4:Summary}.\,---\,We computed a range of properties of the dressed-quark core of the proton's radial excitation and in all cases found they provide an excellent understanding and description of data on the proton-Roper transition and related quantities derived using dynamical coupled channels models.  Our analysis is based on a sophisticated continuum framework for the three-quark bound-state problem; all elements employed possess an unambiguous link with analogous quantities in QCD; and no parameters were varied in order to achieve success.  Moreover, no material improvement in these results can be envisaged before either the novel spectral function methods introduced in Ref.\,\cite{Chang:2013pq} have been extended and applied to the entire complex of nucleon, Delta and Roper properties that are unified herein or numerical simulations of lattice-regularised QCD become capable of reaching the same breadth of application and accuracy.

On the strength of these results and remarks we conclude that the observed Roper resonance is at heart the nucleon's first radial excitation and consists of a well-defined dressed-quark core augmented by a meson cloud that reduces its (Breit-Wigner) mass by approximately 20\%.  Our analysis shows that a meson-cloud obscures the dressed-quark core from long-wavelength probes; but that it is revealed to probes with $Q^2 \gtrsim 3 m_N^2$.  This feature is typical of nucleon-resonance transitions; and hence measurements of resonance electroproduction on this domain can serve as an incisive probe of quark-gluon dynamics within the Standard Model, assisting greatly in mapping the evolution between the nonperturbative and perturbative domains of QCD.

\smallskip

%
We are grateful for insightful comments and suggestions from
R.~Gothe,
T.-S.\,H.~Lee,
V.~Mokeev
and T.~Sato.
J.\,Segovia acknowledges financial support from a postdoctoral IUFFyM contract at the
Universidad de Salamanca.
Work also supported by:
S\~ao Paulo Research Foundation (FAPESP) under grant nos.\ 2012/03275-8 and 2013/16088-4; %
CNPq fellowship nos. 301190/2014-3 and 458371/2014-9;
Patrimonio Aut\'onomo Fondo Nacional de Financiamiento para la Ciencia, la Tecnolog\'ia y la Innovaci\'on, Francisco Jos\'e de Caldas and Sostenibilidad-UDEA 2014-2015;
U.S.\ Department of Energy, Office of Science, Office of Nuclear Physics, under contract no.~DE-AC02-06CH11357;
the National Natural Science Foundation of China (grant nos.\ 11275097
and 11475085); the National Basic Research Programme of China (grant no.\ 2012CB921504);
and the Fundamental Research Funds for the Central Universities Programme of China (grant no.\ WK2030040050).
%


\end{document}